\documentclass[conference]{IEEEtran}
	\IEEEoverridecommandlockouts

  \usepackage{fancyhdr,lipsum}

    \fancypagestyle{mahmood}{%
   
   \fancyhead[C]{As seen in IEEE HONET 2023, Boca Raton, FL, USA.}
    }%

    \makeatletter
    \let\ps@IEEEtitlepagestyle\ps@mahmood
    \makeatother
	\usepackage{cite}
	\usepackage{amsmath,amssymb,amsfonts}
	\usepackage[noend]{algorithmic}
	\usepackage[dvisvgm, usenames, dvipsnames]{color}
	\usepackage{graphicx}
	\usepackage{textcomp}
	\usepackage{xcolor}
	\usepackage{verbatim}
	\usepackage{afterpage}
	\usepackage{ragged2e}
	\usepackage{comment}
	\usepackage[justification=centering]{caption}

    \def\BibTeX{{\rm B\kern-.05em{\sc i\kern-.025em b}\kern-.08em T\kern-.1667em\lower.7ex\hbox{E}\kern-.125emX}}
	\usepackage[margin=1in,footskip=0.25in]{geometry}
\begin{document}
	
		\title{Ransomware Detection Using Federated Learning with Imbalanced Datasets}
	\author{\IEEEauthorblockN{A. Vehabovic$^{1}$, H. Zanddizari$^{1}$, N. Ghani${^1}$, G. Javidi$^{1}$, S. Uluagac$^{2}$, \\ M. Rahouti$^{3}$, E. Bou-Harb$^{4}$,  M. Safaei Pour$^{5}$\\
	\textit{$^{1}$University of South Florida, $^{2}$Florida International University, $^{3}$Fordham University, $^{4}$Louisiana State University, $^{5}$San Diego State University}}
	}
	\maketitle

\maketitle

\begin{abstract}
Ransomware is a type of malware which encrypts user data and extorts payments in return for the decryption keys.  This cyberthreat is one of the most serious challenges facing organizations today and has already caused immense financial damage. As a result, many researchers have been developing techniques to counter ransomware. Recently, the \textit{federated learning} (FL) approach has also been applied for ransomware analysis, allowing corporations to achieve scalable, effective detection and attribution without having to share their private data. However, in reality there is much variation in the quantity and composition of ransomware data collected across multiple FL client sites/regions. This imbalance will inevitably degrade the effectiveness of any defense mechanisms. To address this concern, a modified FL scheme is proposed using a weighted cross-entropy loss function approach to mitigate dataset imbalance. A detailed performance evaluation study is then presented for the case of static analysis using the latest Windows-based ransomware families. The findings confirm improved ML classifier performance for a highly imbalanced dataset.
\end{abstract}
%
%\vspace{-0.10in}
\begin{IEEEkeywords}
Cybersecurity, malware, ransomware analysis, federated learning, data imbalance
\end{IEEEkeywords}

%\vspace{-0.10in}
\section{Introduction}
The ransomware threat has been steadily expanding in recent years, with more sophisticated families and variants emerging on a regular basis, see surveys in \cite{berrueta2019}, \cite{mouss2021}, \cite{vehabovic2022}. This malware uses a range of tactics to infect end-user machines and then proceeds to encrypt important data/files. Criminal affiliates and cyber gangs have even commercialized this malware through a range of \textit{ransomware-as-a-service} (RaaS) offerings, further expanding the threat surface. As a result, there have been many high-profile ransomware incidents that have caused much financial and reputational damage \cite{vehabovic2022},\cite{kapoor2022}.

To address this critical challenge, researchers have developed many different ransomware analysis schemes, and these efforts can be classified using a range of taxonomies, e.g., such as static and dynamic methods, network-based or host-based methods, etc \cite{mouss2021}, \cite{vehabovic2022}. By and large, most of these solutions use some type of \textit{machine learning} (ML) approach, i.e., by extracting and pre-processing raw data and using it to train appropriate classifiers for detection and attribution. 

Nevertheless, collecting and pre-processing large amounts of ransomware data and training ML algorithms is a challenging task, i.e., in terms of privacy and scalability. Many organizations are unwilling to share their private internal host and/or network data.  At the same time, it is becoming unfeasible to perform large-scale data collection and ML training at a single site.  Hence the distributed \textit{federated learning} (FL) approach offers much potential here \cite{mcmahan2017}.  Along these lines, a recent study in \cite{vehabovic2023_2} has applied this method to ransomware analysis, with the findings showing very promising results for the latest Windows-based ransomware families. However, this initial study assumes ``balanced'' datasets with an even distribution of malware samples across all client sites. Clearly, real-world settings will not mirror these idealized scenarios. Instead, differing regions and client bases will likely see (and collect) very different types and amounts of ransomware data. This dataset imbalance will inevitably bias the FL training process and lead to degraded ML classifier performance.

In light of the above, this paper extends the work in \cite{vehabovic2023_2} to address the critical dataset imbalance problem in FL-based ransomware analysis. Namely, the proposed solution uses a weighted cross-entropy loss function approach to account for uneven data distribution across client sites. Overall, this paper is organized as follows. First, Section \ref{survey} briefly reviews existing work in ransomware analysis. Next, Section \ref{FL_weighted_entropy} presents the proposed scheme for imbalanced datasets. The ransomware repository is then detailed in Section \ref{performance}, followed by performance results for binary detection and multi-class attribution using static analysis. Conclusions and future work directions are then presented in Section \ref{conclusions}.

\section{Literature Review}
\label{survey}
A diverse sets of schemes have been proposed to tackle the ransomware threat \cite{berrueta2019}-\cite{vehabovic2022}. Some of these methods are more geared towards early (distribution) stages in the ransomware kill-chain, whereas other methods are more latent and focused on post-infection response and even forensic analysis.  A brief overview is presented here.

Static ransomware analysis schemes analyze binary executable files to detect malicious payloads, e.g., via author attribution, code/segment identification, etc \cite{mouss2021}. Researchers have also proposed ML-based techniques for \textit{binary code analysis} (BCA), source code analysis (reverse engineering), and domain prediction \cite{poudyal2018}, \cite{zhang2019}. In addition, some more specialized solutions also generate image-based training data to test binary classifiers \cite{vehabovic2022}. Many of these methods show good detection accuracies in the mid-90\% range. Researchers have also leveraged Windows \textit{portable executable} (PE) format files for feature extraction \cite{kim2016},\cite{vehabovic2023_1}.  For example, the recent work in \cite{vehabovic2023_1} investigates static PE file analysis using an up-to-date ransomware repository, and results show very good detection and attribution (90-95\% range).

Meanwhile, dynamic network-based ransomware analysis schemes examine traffic information such as packet traces, \textit{domain name service} (DNS) queries, and network storage access, etc. These techniques are also very effective at detecting ransomware activity, with detection rates typically varying around 95\% \cite{almash2019}, \cite{roy2021}.  Similarly, dynamic host-based ransomware schemes have also been studied. These solutions monitor local machine activity to detect ransomware, e.g., memory and file operations, \textit{application programmer interface} (API) function calls, \textit{dynamic link library} (DLL) calls, etc. Some of these methods also try to recover locally-stored encryption keys, see \cite{kolodenker2017}.

Despite these contributions, major concerns still remain. Foremost, many existing studies use mixed training datasets containing older (inactive) ransomware families designed for earlier Windows 7/8 systems (mid-2010s). Hence it is imperative to build updated threat repositories using newer ransomware families, specifically those targeting Windows 10/11 systems. The work in \cite{vehabovic2023_1} presents one such recent effort.  From an operational perspective, it is also very difficult to centralize all data collection and ML training functionality to a single site. Namely, privacy concerns will prevent many organizations from sharing their sensitive data externally. Scalability challenges will also arise if raw ransomware datasets become excessively large.

To address these concerns, recent work in \cite{vehabovic2023_2} presents a distributed ransomware analysis solution based upon the \textit{federated learning} (FL) approach. This effort extends the work in \cite{vehabovic2023_1} and evenly partitions the (Windows PE file) training datasets across multiple sites. Localized datasets are then used to train and average multiple NN-based classifiers using the {\tt FedAvg} algorithm \cite{mcmahan2017}. Findings confirm that the FL approach slightly exceeds centralized ML schemes in terms of detection and attribution accuracy. However, this initial study assumes even data distribution, with local sites having the same number of samples of each family. Clearly, ideal balanced scenarios will rarely exist in practice. Instead, the number of samples and their relative proportions will vary widely across regions. For example, many ransomware attacks target specific users or applications \cite{vehabovic2022}. Hence local repositories will exhibit a high degree of \textit{imbalance} and even non-iid characteristics \cite{lwang2021}. These conditions may even induce majority and minority classes in the training data. Expectedly, the latter will have lower classification accuracy, and this will be particularly problematic for new ransomware releases with fewer samples (such as zero-day attacks).  Hence there is a clear need to develop FL ransomware analysis schemes to handle such dataset imbalance.

\begin{figure*}[h]
    \centering
    \includegraphics[width=5.75in, height=2.55in]{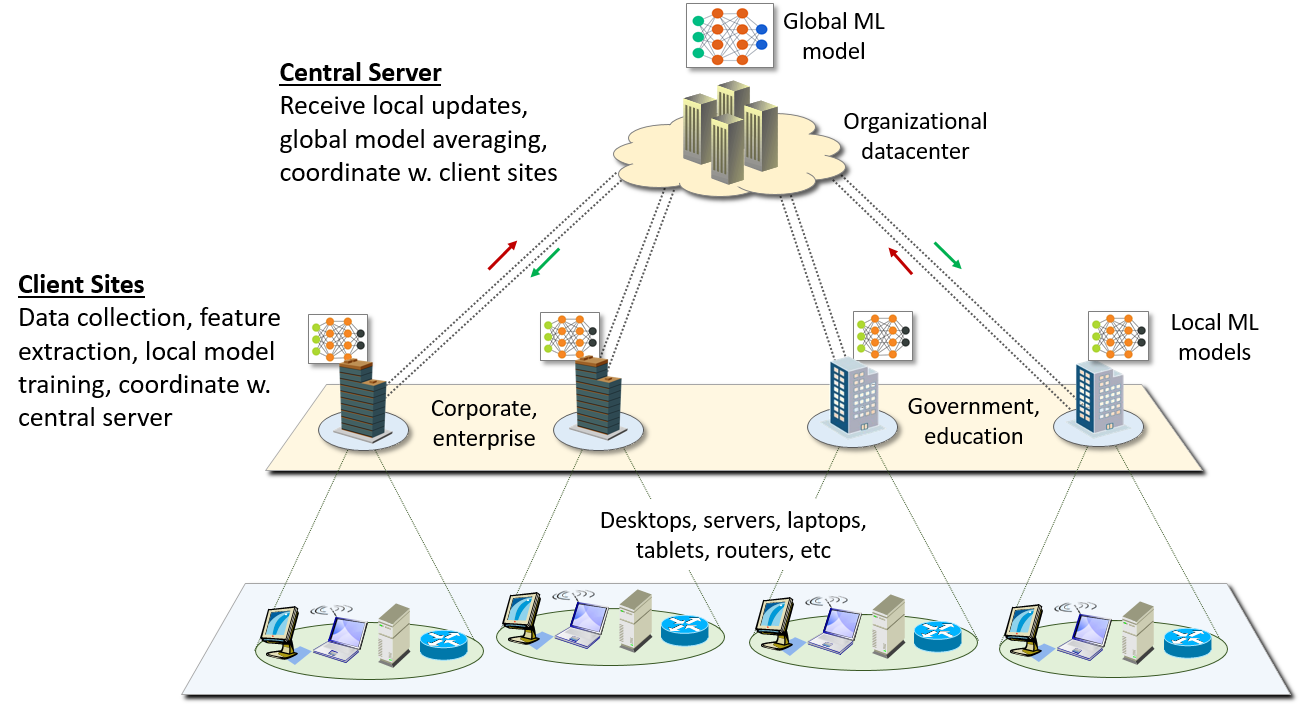}
    \caption{Centralized federated learning approach for ransomware analysis, from \cite{vehabovic2023_2}}
    \label{FL_arch}
\end{figure*}
%
%%%%%%%%%%%%%%%%%%%%%%%%%%%%%%%%%%%%%%%%%%%%%%%%%%%%%%%%%%%%%%
\begin{figure}[!ht]
\line(1,0){225}
\begin{algorithmic}
\small
\STATE \textbf{Input:} Global ML parameters from central server, $\omega$ 
\\
\vspace{0.05in}
\STATE \textit{/* Compute inverse class frequency weights */}
  \STATE Compute weighting vector $\{ \alpha_1, \alpha_2, \cdots, \alpha_Q \} \rightarrow \Pi_K$
\\
\vspace{0.05in}
\STATE Partition local ML dataset ($n_i$ samples) into $B$ batches
\vspace{0.05in}
\STATE \textit{/* Train local ML model over given epochs and batches */}
\FOR{j=1 to $E$}
\FOR{k=1 to $B$} 
  \STATE Train local model over $k$-th batch of data
  \STATE Update loss function w.r.t. class frequencies, $\Pi_Q$  
  \STATE Update local ML model parameters $\omega \rightarrow \omega_i^t$
\ENDFOR
\ENDFOR
\vspace{0.05in}
\STATE \textbf{Output:} Send updated ML parameters, $\omega_i^t$, to central server
\end{algorithmic}
\line(1,0){225}
\caption{Client site $i$ algorithm }
\label{psuedocode_weightedloss_client}
\end{figure}
%%%%%%%%%%%%%%%%%%%%%%%%%%%%%%%%%%%%%%%%%%%%%%%%%%%%%%%%%%%%%%
%%%%%%%%%%%%%%%%%%%%%%%%%%%%%%%%%%%%%%%%%%%%%%%%%%%%%%%%%%%%%%
\begin{figure}[!ht]
\line(1,0){225}
\begin{algorithmic}
\small
\STATE  \textbf{Input:} Initial global ML parameters, $\omega_0$ 
\\
\vspace{0.05in}
\STATE \textit{/* Iterative FL training process over $T$ rounds */} 
\FOR {$t=1$ to $T$}

\vspace{0.05in}
\STATE Select $K$ client sites 
\\
\vspace{0.05in}
\FOR {$i=1$ to $K$}
\STATE Send latest global ML model $\omega$ to $i$-th client site
\ENDFOR
\vspace{0.05in}
\STATE Wait to receive local updates from all $K$ client sites, $\omega_i^t$ 
\\
\vspace{0.025in}
\STATE Average to update global ML model parameters:
\\
\hspace{0.45in} $\omega  \leftarrow \omega^{t+1}=\sum_{i=1}^{K} \frac{n_i}{N} \omega_i^t$
\ENDFOR
\vspace{0.05in}
\STATE \textbf{Output:} Final global ML parameters, $\omega$ \\
\end{algorithmic}
\line(1,0){225}
\caption{Centralized server algorithm, {\tt FedAvg} \cite{mcmahan2017} }
\label{psuedocode_server}
\end{figure}
%%%%%%%%%%%%%%%%%%%%%%%%%%%%%%%%%%%%%%%%%%%%%%%%%%%%%%%%%%%%%%

\section{Federated Learning with Imbalanced Ransomware Datasets}
\label{FL_weighted_entropy}
As noted earlier, the distributed FL framework has recently been applied to ransomware analysis \cite{vehabovic2022}.  This setup comprises of a \textit{central server} communicating with multiple \textit{client sites}, Figure \ref{FL_arch}. Here, the former entity coordinates the overall distributed FL process. Meanwhile, the latter entities are located at trusted organizations with large user bases and perform raw data collection, feature extraction, and local ML model training. Namely, local training is done over $T$ rounds, in which client sites receive ``global'' model parameters from the central server, $\omega$. These models are then trained using local data and the updates sent back to the central server, see psuedocode in Figure \ref{psuedocode_weightedloss_client}. Specifically, client site $i$ has $n_i$ local samples that are split into $B$ batches for training $E$ epochs. Meanwhile, the central server selects a subset of $K$ client sites to receive the latest global model parameters in each round, see psuedocode in Figure \ref{psuedocode_server}. Incoming parameter updates are then averaged in a weighted manner to update the global model, where $\omega_i^t$ is the updated model parameters from client site $i$ in round $t$ ({\tt FedAvg} algorithm \cite{mcmahan2017}).

Now results in \cite{vehabovic2023_2} confirm that FL is very effective for ransomware detection and attribution with perfectly balanced client datasets. However, data imbalance can skew locally-trained models and reduce global accuracy \cite{duan2019}. In particular, uneven data distribution can exist at both the global and local levels in FL, further complicating the challenge. Although anonymizing and sharing data between client sites is an option, few organizations will allow this. Moreover, data sharing violates a key premise of FL privacy. Now data imbalance in ML is not a new problem, and various solutions have been proposed, broadly categorized as data-level, algorithm-level, and hybrid \cite{lwang2021}. The goal is to improve ML classifier performance and closely track results with balanced datasets. For example, data-level methods use sampling and generative augmentation to add artificial samples for minority classes \cite{pouyanfar2018}. Conversely, algorithm-level schemes use revised ML algorithms such as modified networks, updated loss functions, etc \cite{lwang2021}. Meanwhile, hybrid methods use a mix of strategies. However, most of these schemes are not designed for FL, although they can be applied at local client sites.

Now recent efforts have also proposed more specialized FL solutions for imbalanced datasets. For example, \cite{duan2019} presents a self-balancing method using data augmentation and down-sampling techniques at client nodes. However, this method uses intermediate proxy servers and requires clients to share local data distributions. The latter is a key concern as it may compromise some privacy. Alternatively, \cite{lwang2021} proposes a monitoring scheme to estimate class composition in the training data during FL operation. This method requires clients to provide total sample counts and uses a weight change threshold to estimate proportion vectors in each training round.  If sufficient imbalance is detected in a client's data composition, then a new ratio loss function is used to mitigate its impact (based on cross entropy loss). Others have also studied the related problem of non-iid training datasets in FL, i.e., different underlying distributions at the client sites, see \cite{vehabovic2023_2} for references. 

Drawing from the above, a weighted approach is used to scale the regular cross entropy loss function used for training \textit{feedforward neural network} (FNN) classifiers (at the client sites). Consider the regular cross entropy loss function for a multi-class problem with $Q$ classes:
\begin{equation}
    L_{CE}=- \sum_{i=1}^{Q} t_i {\tt log}(p_i)
    \label{cross_entropy_loss_fn}
\end{equation}
where $t_i$ is the ground truth label for class $i$, and $p_i$ is the softmax probability for class $i$. To address dataset testing imbalance, a weighting vector, $\Pi=\{ \alpha_1, \alpha_2, \cdots, \alpha_Q \}$, is introduced to scale the classes based upon the proportion of training data for each, i.e., where $\alpha_j$ is the weight for each individual class $j$. The overall strategy here is to assign larger weights to classes with more scarce data, and assign smaller weights to those with more representation. Now the individual values for $\alpha_j$ are determined as follows: 
\begin{equation}
    \alpha_j = \Big( \frac{\tt n_i^j}{{\tt max} ({\tt{Z^T)}}} \Big) ^{-1}
    \label{ch6_alpha_i}
\end{equation}
where $n_i^j$ is the number of training samples of class $j$ at client site $i$, and $Z_T$ is an array representing all training values of $n_i^j$ at a given client site $i$. This equation is a variation of a technique known as inverse class frequency weighting, where higher frequency classes are scaled using smaller weights and lower frequency classes are scaled using larger weights. These revisions for weighted cross entropy loss are also shown in the client site psuedocode in Figure \ref{psuedocode_weightedloss_client} (versus \cite{vehabovic2023_2}).

\begin{figure*}[h]
    \centering
    \includegraphics[width=5.0in, height=3.30in]{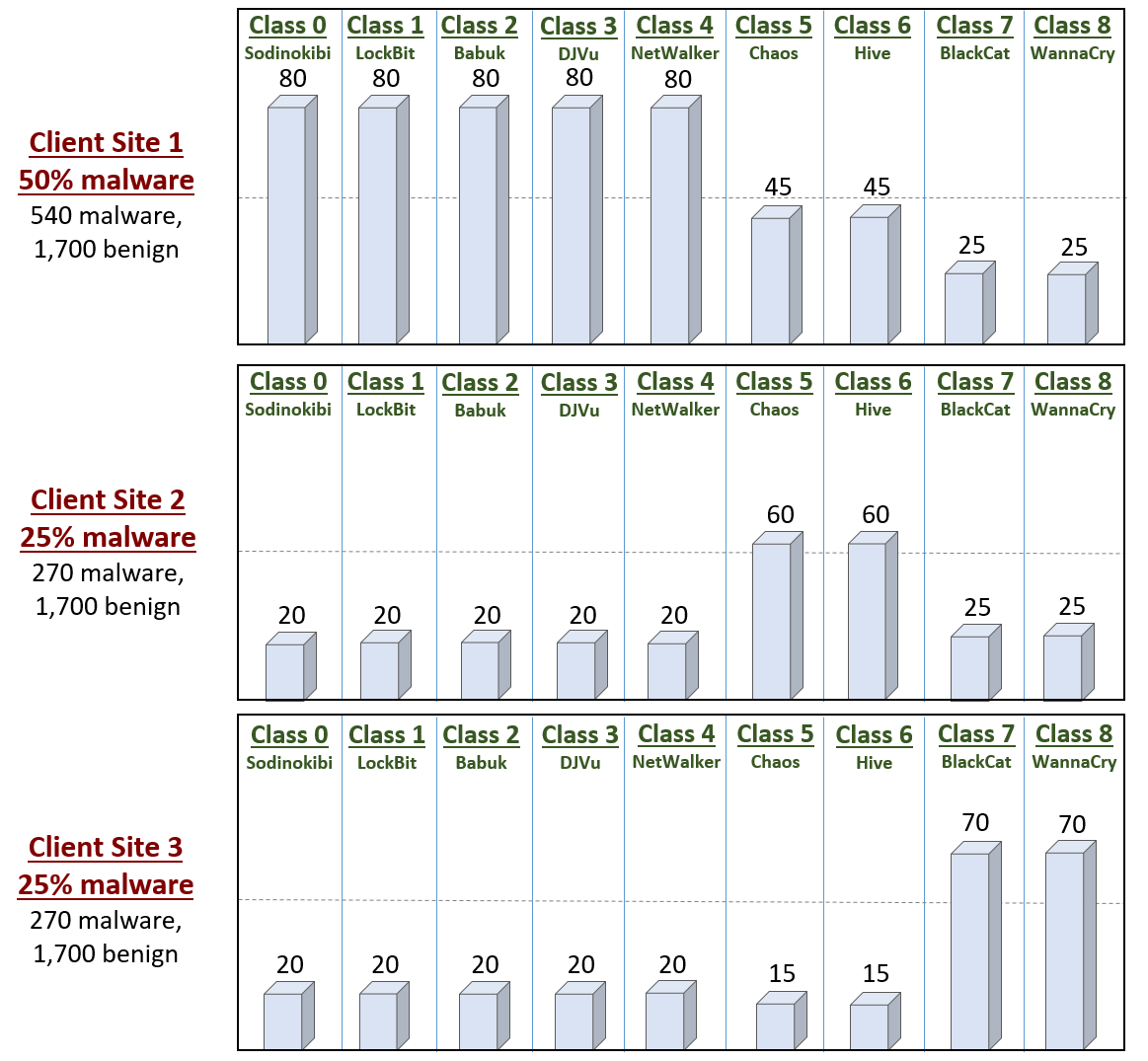}
    \caption{Imbalanced ransomware dataset for FL setup with $K$=3 client sites (using FNN classifiers)}
    \label{dataset_imbalanced}
\end{figure*}

\begin{table}[h]
	\centering
\caption{Empirical dataset (9 families and benign)}
\label{table_dataset}
\begin{tabular}{|l|c|c|c|} 
\hline
\textbf{ Family } & \textbf{Samples} & \textbf{Avg. Size}  & \textbf{Avg. PE File} \\ \hline \hline
Babuk/Babyk & 140 & 0.19 MB & 32.68 KB \\ \hline
BlackCat & 120 & 3.91 MB & 1,147 KB \\ \hline
Chaos & 140 & 0.49 MB & 35.2 KB \\ \hline
DJVu (STOP) & 140 & 0.71 MB & 66.2 KB \\ \hline
Hive & 140 & 3.51 MB & 403.9 KB \\ \hline
LockBit & 140 & 1.30 MB & 171.5 KB \\ \hline
Netwalker & 140 & 0.26 MB & 35.72 KB \\ \hline
Sodinokibi & 140 & 0.30 MB & 50.89 KB \\ \hline
WannaCry & 140 & 7.62 MB & 21.83 KB \\ \hline
\hline
Benign apps & 2,000 & 26.86 MB & 155.88 KB \\ \hline
\end{tabular}  
\end{table}

\section{Performance Evaluation}
\label{performance}
The performance of the weighted cross entropy loss scheme is tested for both the ransomware detection and attribution problems. An up-to-data repository is used for evaluation purposes here, as per recent studies in \cite{vehabovic2023_2},\cite{vehabovic2023_1}.  Specifically, this dataset contains binary executables of some of latest ransomware families, as per the IBM X-Force Threat Intelligence Index. This collection includes Babuk/Babyk, BlackCat, Chaos, DJVu/STOP, Hive, LockBit, Netwalker, Sodinokibi/REvil, and WannaCry (9 families in total).

Now given the dynamic nature of the ransomware threat, it is difficult to collect a large number of samples for each family. Hence, only 140 executable samples are collected for each family from various open repositories, e.g., such as \textit{MalwareBazar}, \textit{Triage}, \textit{VirusShare}, and \textit{VirusTotal}. This gives a total of 1,260 malware samples. In addition, 2,000 benign Windows applications are also collected to build a benign training class, i.e., 10 total classes.  These raw binary executables are then pre-processed to build the ML training/testing dataset.  In particular, 15 specific parameters are extracted from the Windows PE format files to generate the requisite feature vectors. Note that these file types are organized in sections (with headers) and contain crucial metadata to support program execution. For further details on the extracted feature sets, please refer to \cite{vehabovic2023_1}.

Now FL evaluation is done for the case of $K=3$ distributed client sites.  Accordingly, the above dataset is split in a balanced (even) and imbalanced (uneven) manner using a 85/15\% partitioning between training/testing data.  Namely, for balanced dataset partitioning, 20 random samples of each ransomware family are selected for testing, and the remaining 120 samples are distributed across the 3 client sites for training, i.e., 40 samples each (only 360 malware samples per client site). Conversely, the benign samples are not partitioned between the client sites. The reason here is that regular applications downloads will exceed ransomware downloads in many real-world settings, and there may also be a strong overlap between application downloads across sites.  Hence it is feasible and realistic to use the same (larger) set of benign samples across all client sites. Hence the benign samples are also split in a 85/15\% manner, with 300 randomized samples selected for testing, and the remaining 1,700 used for training. 

Meanwhile, the imbalanced dataset is also designed using an 85/15\% partitioning. Again, 20 random samples from each family are reserved for testing, whereas the remaining 120 samples are used for training (1,080 total). However, the latter samples are now distributed in an uneven manner between the 3 client sites, as shown in Figure \ref{dataset_imbalanced}.  This partitioning generates a notable imbalance between the number of training samples at each client site and the relative percentage of families at each client site. For example, client site 1 has 50\% of the training data (540 samples), whereas client sites 2 and 3 have 25\% (270 samples each). Furthermore, classes 0-4 (Sodinokibi, LockBit, Babuk/Babyk, DJVu, NetWalker) represent the majority classes at client site 1. Conversely, classes 5 and 6 (Chaos, Hive) represent the majority classes at client site 2, and classes 7 and 8 (BlackCat, WannaCry) represent the majority classes at client site 3, Figure \ref{dataset_imbalanced}. To further quantify this imbalance, the \textit{imbalance ratio} from \cite{lwang2021} is also computed for each client site $i$ as follows:
\begin{equation}
    \gamma_i = \frac{{\tt max}_j ({n_i^j}) }{{\tt min}_j ({n_i^j})}
    \label{imbalance_ratio}
\end{equation}
where $n_i^j$ is the number of samples of class $j$ at client site $i$ (as defined in Eq. \ref{ch6_alpha_i}).  Hence values closer to unity imply more even (balanced) data distribution at a client site. Based on the above, the imbalance ratios for multi-class attribution (between the ransomware classes) are $\gamma_1$=3.2, $\gamma_2$=3.0, and $\gamma_3$=4.7, respectively, for the three client sites. If also taking into account the larger benign class (1,700 samples), the corresponding imbalance ratios are much higher, i.e., $\gamma_1$=65, $\gamma_2$=85, and $\gamma_3$=113.33.  Meanwhile, for the simpler binary detection problem, the imbalance ratios between the two classes are $\gamma_1$=3.14, $\gamma_2$=6.29, and $\gamma_3$=6.29. Note that \cite{lwang2021} also defines a global imbalance ratio, i.e., give by the total number of samples of the largest majority class divided by the total number of samples of the smallest minority class. However, since there are 120 testing samples for each ransomware family here, there are no majority or minority classes at the global level here, i.e., imbalance only occurs between the client sites.

The proposed FL approach is now analyzed using FNN classifiers (using weighted cross entropy loss). All evaluation is done using the {\tt Keras} and {\tt TensorFlow} toolkits, as well as {\tt Pandas} and {\tt Sklearn}. Furthermore, results for each test scenario are averaged over 100 randomized trials. Results for the binary ransomware detection problem are presented first. Here, all ransomware training samples are combined to generate a single malware class for training purposes, i.e., 540 samples at client site 1 and 270 samples each at client sites 2 and 3 (Figure \ref{dataset_imbalanced}). The average values for accuracy, precision, recall, and F1 score are summarized in Table \ref{FL_binary_imbalanced_table}. In particular, results are presented for three scenarios, i.e., baseline FL scheme with balanced dataset (and regular cross entropy loss function), baseline FL scheme with imbalanced dataset (and regular cross entropy loss function), and proposed FL scheme with imbalanced dataset (and weighted cross entropy loss function). These findings indicate that the relative improvement between the global and local client models (due to global FL model averaging) is much smaller with the imbalanced dataset, i.e., under 2.5\% for most metrics. There is also a small, but consistent, performance degradation as compared to the balanced dataset results, albeit mostly under 2\%. However, the weighted cross entropy loss scheme does better than the regular weighting scheme, yielding an average accuracy within 0.5\% of that with the balanced dataset (and most metrics approaching 95\% range).

Results for the multi-class attribution problem are also presented in Table \ref{FL_multi_imbalanced_table} ($K$=3). Again, all three scenarios are evaluated here, i.e., baseline FL scheme using balanced dataset, baseline FL scheme using imbalanced dataset, and proposed FL scheme using imbalanced dataset. Overall, these findings show good performance improvement with global model averaging (in the FL schemes) for the imbalanced dataset, albeit the relative differences between the global and local models are smaller with the weighted cross entropy loss function approach (about 2-3\% for most metrics).  However, unlike the binary detection results in Table \ref{FL_binary_imbalanced_table}, there are more significant performance declines versus the results with the balanced dataset. In particular, since there much fewer samples of some minority classes at certain client sites, attribution is much more difficult (versus simple binary detection with larger aggregated classes). For example, average accuracy and F1 scores are 8-10\% lower than the perfectly balanced scenarios, whereas the average precision and recall rates are 2-6\% lower.  Again, the weighted cross entropy scheme gives slightly better accuracy than the regular scheme, i.e., about 2\% higher. Although one can argue that the imbalance ratios for the uneven distribution case are very high to begin with, these findings indicate a further need to develop new schemes for imbalanced scenarios.

%%%%%%%%%%%%%%%%%%%%%%%%%%%%%%%%%%%%%%%%%%%%%%%%%%%%%%%%%%%%%%%%
\begin{table}[h]
	\centering
\caption{Binary detection ($K$=3 client sites)}
\label{FL_binary_imbalanced_table}
\small
\begin{tabular}{|c|c|c|c|c|} 
\hline
\multicolumn{5}{|c|}{\textbf{Baseline {\tt FedAvg} (balanced dataset) }}\\
\hline
\textbf{ } & \textbf{ Acc. } & \textbf{ Prec. }  & \textbf{ Recall } & \textbf{ F1 } \\ \hline 
{\textbf{Global}} & \textbf{95.08\%} & \textbf{95.10\%} & \textbf{94.43}\% & \textbf{95.25}\% \\ \hline
Client 1 & 89.18\% & 89.18\% & 89.25\% & 88.36\% \\ \hline
Client 2 & 90.03\% & 89.32\% & 90.17\% & 89.28\% \\ \hline
Client 3 & 90.23\% & 89.44\% & 90.23\% & 89.40\% \\ \hline \hline
%
%\multicolumn{5}{|l|}{ } \\ 
\multicolumn{5}{|c|}{\textbf{Baseline {\tt FedAvg} (imbalanced dataset)}}\\
\hline
\textbf{ } & \textbf{ Acc. } & \textbf{ Prec. }  & \textbf{ Recall } & \textbf{ F1 } \\ \hline 
{\textbf{Global}} & \textbf{ 93.86\%} & \textbf{ 93.94\%} & \textbf{ 93.86\%} & \textbf{ 93.86\%} \\ \hline  % 4 Epochs , Communication rounds = 50
Client 1 &  92.95\% &  92.90\% &  92.94\% &  92.93\% \\ \hline
Client 2 &  92.58\% &  92.71\% &  92.58\% &  92.57\% \\ \hline
Client 1 &  92.72\% &  92.58\% &  92.71\% &  92.67\% \\ \hline \hline
%
%\multicolumn{5}{|l|}{ } \\ 
\multicolumn{5}{|c|}{\textbf{Weighted cross entropy loss (imbalanced dataset)}}\\
\hline
\textbf{ } & \textbf{ Acc. } & \textbf{ Prec. }  & \textbf{ Recall } & \textbf{ F1 } \\ \hline 
{ \textbf{Global} } & \textbf{ 94.67\%} & \textbf{ 94.70\%} & \textbf{ 94.62\%} & \textbf{ 94.66\%} \\ \hline  % 4 Epochs , Communication rounds = 50
Client 1 &  92.70\% &  92.97\% &  92.69\% &  92.68\% \\ \hline
Client 2 &  92.64\% &  92.90\% &  92.64\% &  92.63\% \\ \hline
Client 1 &  92.03\% &  92.36\% &  92.03\% &  92.01\% \\ \hline \hline
\end{tabular}  
\end{table}
%%%%%%%%%%%%%%%%%%%%%%%%%%%%%%%%%%%%%%%%%%%%%%%%%%%%%%%%%%%%%%%%
%
%%%%%%%%%%%%%%%%%%%%%%%%%%%%%%%%%%%%%%%%%%%%%%%%%%%%%%%%%%%%%%%%
\begin{table}[h]
	\centering
\caption{Multi-class attribution ($K$=3 client sites)}
\label{FL_multi_imbalanced_table}
\small
\begin{tabular}{|c|c|c|c|c|}
\hline
\multicolumn{5}{|c|}{\textbf{Baseline {\tt FedAvg} (balanced dataset) }}\\
\hline
 & \textbf{ Acc. } & \textbf{ Prec. }  & \textbf{ Recall } & \textbf{ F1 } \\ \hline 
{ \textbf{Global} } & \textbf{92.11\%} & \textbf{87.40\%} & \textbf{86.57\%} & \textbf{91.90\%} \\ \hline Client 1 & 86.02\% & 77.86\% & 75.64\% & 85.33\% \\ \hline
Client 2 & 82.90\% & 73.89\% & 74.56\% & 82.13\% \\ \hline
Client 3 & 81.47\% & 73.55\% & 75.95\% & 81.49\% \\ \hline 
% & \multicolumn{2}{|c|}{ Global RDR: 93.04\% } & \multicolumn{2}{|c|}{ Global BDR: 95.86\% } \\ \hline
%
\hline
\multicolumn{5}{|c|}{\textbf{Baseline {\tt FedAvg} (imbalanced dataset) }}\\
\hline
 & \textbf{ Acc. } & \textbf{ Prec. }  & \textbf{ Recall } & \textbf{ F1 } \\ \hline 
{ \textbf{Global} } & \textbf{82.55\%} & \textbf{85.17\%} & \textbf{82.55\%} & \textbf{82.80\%} \\ \hline
Client 1 & 81.70\% & 84.17\% & 81.7\% & 81.73\% \\ \hline
Client 2 & 77.85\% & 82.71\% & 77.85\% & 77.80\% \\ \hline
Client 3 & 76.95\% & 81.13\% & 76.95\% & 76.40\% \\ \hline
% & \multicolumn{2}{|c|}{ Global RDR:  87.00\% } & \multicolumn{2}{|c|}{ Global BDR: 93.00\% } \\ \hline \hline
%
\multicolumn{5}{|c|}{\textbf{Weighted cross entropy loss (imbalanced dataset) }}\\
\hline
 & \textbf{ Acc. } & \textbf{ Prec. }  & \textbf{ Recall } & \textbf{ F1 Score } \\ \hline 
{\textbf{Global} } & \textbf{84.15\%} & \textbf{85.35\%} & \textbf{82.05\%} & \textbf{82.35\%} \\ \hline  
Client 1 & 83.35\% & 81.99\% & 82.05\% & 82.35\% \\ \hline
Client 2 & 82.15\% & 83.82\% & 79.85\% & 80.29\% \\ \hline
Client 3 & 81.85\% & 83.56\% & 79.90\% & 80.09\% \\ \hline
%
%& \multicolumn{2}{|c|}{ Global RDR:  89.71\% } & \multicolumn{2}{|c|}{ Global BDR: 96.00\% } \\ \hline \hline
%
\end{tabular}  
\end{table}
%%%%%%%%%%%%%%%%%%%%%%%%%%%%%%%%%%%%%%%%%%%%%%%%%%%%%%%%%%%%%%%%
%
\section{Conclusions}
\label{conclusions}
The \textit{federated learning} (FL) framework provides improved privacy and scalability support for large \textit{machine learning} (ML) problems. Here, multiple client sites are used to train classifier algorithms using local data, and the parameters averaged to build a more accurate global model. Now FL has been applied to a wide range of applications, including a recent study on ransomware detection and attribution. However, dataset imbalance can limit the effectiveness of this method. Hence this study presents a modified FL learning scheme for ransomware analysis which uses a weighted cross entropy loss function to train regular \textit{feedforward neural network} (FNN) classifiers at local client sites. This solution is evaluated using an up-to-date repository containing the latest ransomware threats for the case of static analysis with Windows \textit{portable executable} (PE) format file data. Results confirm that the proposed scheme can help mitigate the effects of dataset imbalance, particularly for the binary ransomware detection problem.

\section{Acknowledgements}
This work has been supported in part by Cyber Florida.  The authors are very grateful for this support.

% References
\bibliographystyle{IEEEtran}
\bibliography{references.bib}

\end{document}